N-terminal domain Increases Activation of Elephant Shark Glucocorticoid and Mineralocorticoid Receptors

Yoshinao Katsu[1, 2], Islam MD Shariful[1], Xiaozhi Lin[1], Wataru Takagi[3], Hiroshi Urushitani[4], Satomi Kohno[5], Susumu Hyodo[3], Michael E. Baker[6, *]

[1]Graduate School of Life Science, Hokkaido University, Sapporo, Japan
[2]Faculty of Science, Hokkaido University, Sapporo, Japan
[3]Laboratory of Physiology, Atmosphere and Ocean Research Institute, University of Tokyo, Chiba, Japan
[4]Department of Food and Nutrition, The University of Aizu, Junior College Division, Fukushima, Japan
[5]Department of Biology, St. Cloud State University, St. Cloud, MN, USA
[6]Department of Medicine, University of California, San Diego, CA, USA

*Correspondence: mbaker@ucsd.edu

**Abstract**. Cortisol, corticosterone and aldosterone activate full-length glucocorticoid receptor (GR) from elephant shark, a cartilaginous fish belonging to the oldest group of jawed vertebrates. Activation by aldosterone a mineralocorticoid, indicates partial divergence of elephant shark GR from the MR. Progesterone activates elephant shark MR, but not elephant shark GR. Progesterone inhibits steroid binding to elephant shark GR, but not to human GR. Deletion of the N-terminal domain (NTD) from elephant shark GR (Truncated GR) reduced the response to corticosteroids, while truncated and full-length elephant shark MR had similar responses to corticosteroids. Chimeras of elephant shark GR NTD fused to MR DBD+LBD had increased activation by corticosteroids and progesterone compared to full-length elephant shark MR. Elephant shark MR NTD fused to GR DBD+LBD had similar activation as full-length elephant shark MR, indicating that activation of human GR by the NTD evolved early in GR divergence from the MR.

Running title: Evolution of steroid specificity of elephant shark glucocorticoid receptor.

Key words: elephant shark GR, glucocorticoid receptor evolution, allosteric regulation, corticosteroids,

**Introduction**

Glucocorticoids, such as cortisol and corticosterone, have diverse physiological activities in humans and other terrestrial vertebrates, including regulating glucose metabolism, cognition, immunity and the response to stress (*1-5*). The physiological actions of glucocorticoids are mediated by the glucocorticoid receptor (GR), which belongs to the nuclear receptor family, a diverse group of transcription factors that also contains receptors for mineralocorticoids (MR), progestins (PR) androgens (AR) and estrogens (ER) (*6-8*).

The GR is closely related to the MR (*9*); phylogenetic analysis indicates that the GR and MR evolved from a corticoid receptor (CR) that evolved in a jawless vertebrate that was an ancestor of modern lamprey and hagfish (*7, 10-13*). Distinct orthologs of human GR and human MR first appear in cartilaginous fishes, the oldest group of extant jawed vertebrates (gnathostomes) that diverged from bony vertebrates about 450 million years ago (*14*). Since the emergence of the GR and MR in cartilaginous fishes, these steroid receptors have diverged to respond to different corticosteroids. For example, human MR is activated by aldosterone, the physiological mineralocorticoid in humans, at a concentration at least ten-fold lower than by cortisol (*15-19*). However, aldosterone has little activity for human GR, which is activated by cortisol (Figure 1) (*10, 13, 15, 17, 20*).

The multi-domain structure of the GR is important in regulating steroid activation of the GR. Like other steroid receptors, the GR consists of an N-terminal domain (NTD) (domains A and B), a central DNA-binding domain (DBD) (domain C), a hinge domain (D) and a C-terminal ligand-binding domain (LBD) (domain E) (*5, 8, 17, 21, 22*) (Figure 2). Allosteric interactions between the NTD and LBD increase corticosteroid activation of GR in humans, other terrestrial vertebrates (*17, 22-28*) and ray-finned fish (*22, 29*). For example, neither cortisol nor corticosterone activate truncated human GR in which the NTD is deleted (*17, 22*). It is not known when the strong dependence of vertebrate GR on the NTD for activation of gene transcription evolved.

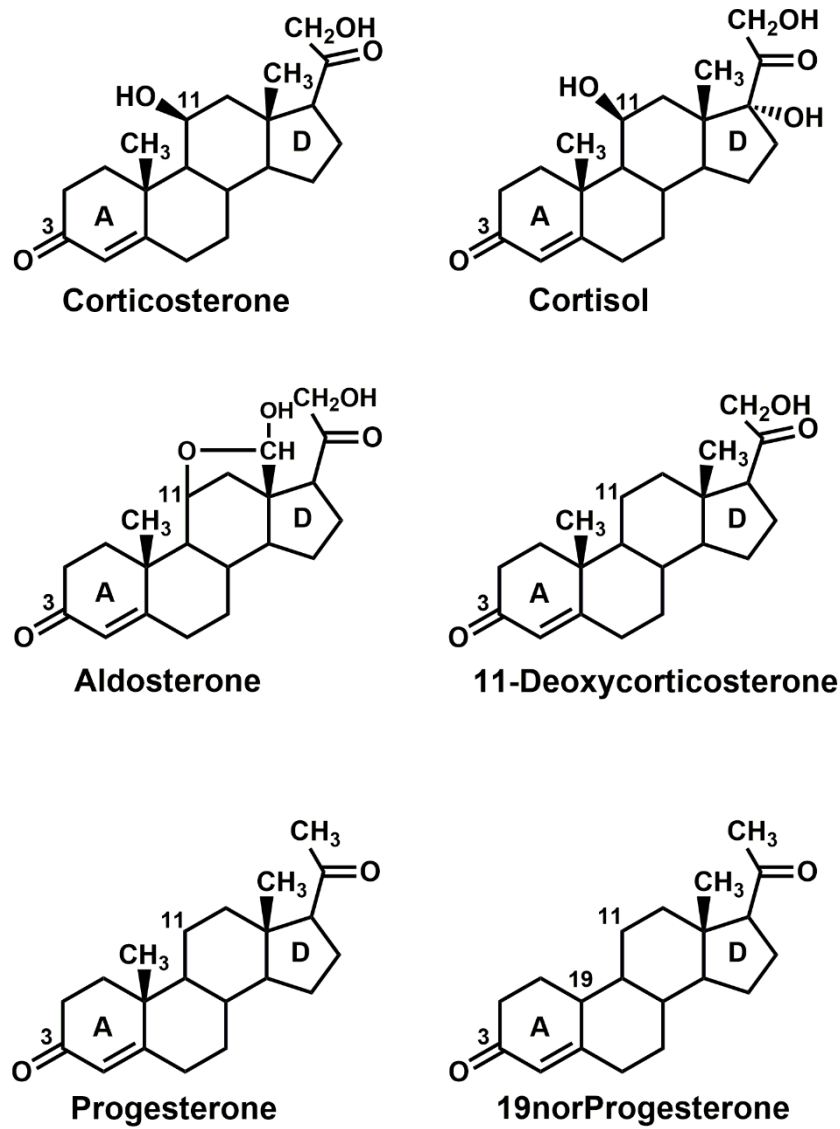

**Fig. 1. Structures of Corticosteroids and Progesterone.** Cortisol and corticosterone are physiological glucocorticoids in terrestrial vertebrates and ray-finned fish (*7, 10, 30*). Aldosterone, 11-deoxycorticosterone are physiological mineralocorticoids (*10, 13, 31, 32*). Progesterone is female reproductive steroid that also is important in male physiology (*33, 34*).

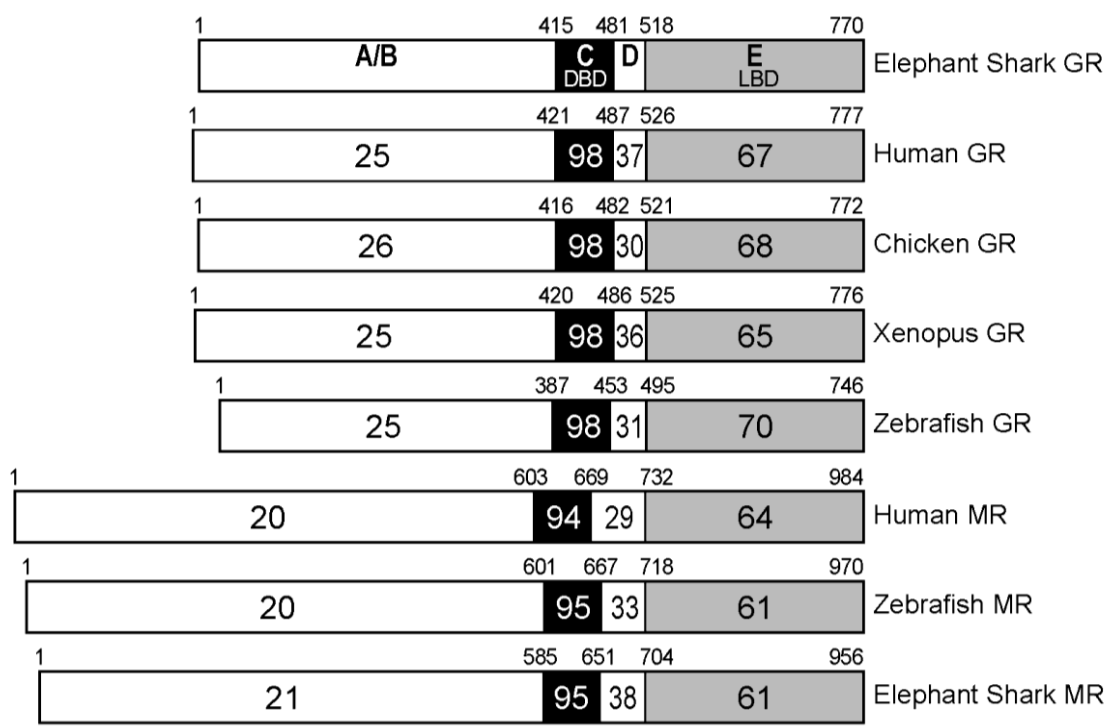

**Fig.2. Comparison of domains in elephant shark GR with vertebrate GRs and MRs.** GRs from elephant shark, zebrafish, *X. laevis*, chicken and humans and MRs from elephant shark, humans and zebrafish are compared. The functional A/B domain to E domains are schematically represented with the numbers of amino acid residues and the percentage of amino acid identity depicted.

To investigate corticosteroid signaling via the GR, in the light of the evolution (*35*), we decided to study activation of full-length and truncated elephant shark (*Callorhinchus milii*) GR by a panel of corticosteroids, which were used in a previous study of elephant shark MR (*36*). Elephant shark is an excellent system for evolutionary studies because in addition to the key phylogenetic position of elephant shark in vertebrate evolution, as a precursor for ray-finned fish and terrestrial vertebrates, genomic analyses reveal that elephant shark genes are evolving slowly, which makes their genes windows into the past (*14*). Thus, characterization of steroid activation of elephant shark GR permits analyses relevant to early events in the divergence of the GR and its MR paralog.

Two questions arise regarding the evolution of glucocorticoid activation of the GR. First, what were the active glucocorticoids early in the evolutionary divergence of the GR from its common ancestor with the MR? That is, what is the response of elephant shark GR to physiological glucocorticoids (cortisol, corticosterone) and mineralocorticoids (aldosterone, 11-deoxycorticosterone), as well as progesterone, all of which activate elephant shark MR (*36*)?    Interestingly, aldosterone, which is not found in cartilaginous fishes (*11, 37*), has a half maximal response (EC50) of 11 nM for truncated skate GR in which the NTD is deleted (*37*).   The EC50 of 11 nM of aldosterone for skate GR is lower than the EC50s of 139 nM and 58 nM, respectively, for cortisol and corticosterone for skate GR (*37*).   However, because an analysis of corticosteroid activation of a full-length cartilaginous fish GR has not been reported, the identity of physiological glucocorticoids in cartilaginous fish is not known.

The absence of data on steroid activation of a full-length cartilaginous fish GR, raises a second question: What is the role of the NTD in corticosteroid activation of the elephant shark GR, and how does it compare to the role of the NTD in human GR (*17, 22, 24*) and elephant shark MR?

As reported here, we find that full-length elephant shark GR is activated by cortisol and corticosterone, which activate human GR (*25*).   Unexpectedly, aldosterone and 11-deoxycorticosterone, two mineralocorticoids, also activated elephant shark GR. These two mineralocorticoids have little activity for human GR (*15, 19, 25*), which is selective for cortisol and corticosterone (*25*).   Elephant shark GR is not activated by either progesterone or 19norprogesterone, in contrast to elephant shark MR, which also is activated by cortisol and corticosterone (*36*).   However, we find that progesterone and 19norprogesterone inhibit corticosterone activation of elephant GR, while neither progestin inhibits cortisol activation of human GR.

We find that activation by corticosteroids of truncated elephant shark GR, in which the NTD is deleted, was less than 10% of that of full-length GR, in contrast to truncated elephant shark MR, in which deletion of the NTD did not have a major effect on corticosteroid activation (*36*).   To investigate specificity of the NTD for activation of the GR and MR, we studied chimeras of elephant shark GR and MR in which the GR NTD was fused to elephant shark MR DBD-LBD and the MR NTD was fused to GR DBD-LBD.   Activation by cortisol and corticosterone of elephant shark GR NTD fused to MR DBD-LBD increased by over 10-fold, compared to full length elephant shark MR.   Interestingly, the chimera of GR NTD MR DBD-LBD had increased activation by progesterone and 19norprogestrone.   In contrast, corticosteroid activation of the MR NTD fused to GR DBD-LBD was reduced by over 90%.   These data indicate allosteric regulation by the NTD in elephant shark GR evolved after the divergence of the GR from its common ancestor with the MR.

**Results**
**Functional domains on elephant shark GR and other vertebrate GRs.**

In Figure 2, we compare the functional domains of elephant shark GR to domains in selected vertebrate GRs (human, chicken, *Xenopus*, zebrafish) and MRs (human, zebrafish and elephant shark).   Elephant shark GR and human GR have 98% and 67% identity in DBD and LBD, respectively.   The DBD and LBD in elephant shark GR is conserved in other GRs.   The A, B and D domains of elephant shark GR are much less conserved with other GRs.   These GRs and elephant shark GR have over 95% and 65% identity in the DBD and LBD, respectively.   Elephant shark GR and mR have good conservation in their D and E domains, with 95% and 61% identity, respectively.

**Effect of corticosteroids and progesterone on transcriptional activation of full-length elephant shark GR and MR.**

Activation of full-length elephant shark GR by $10^{-8}$ M cortisol, corticosterone, 11-deoxycorticosterone, aldosterone, progesterone and 19norProgesterone is shown in Figure 3A.   For comparison activation of elephant shark MR by these steroids is shown in Figure 3B.   Cortisol, corticosterone, 11-deoxycorticosterone, and aldosterone activated full-length elephant shark GR (Figure 3A) and MR (Figure 3B).   However, activation of full-length elephant shark GR by corticosteroids at $10^{-8}$ M was over 10-fold higher than for full-length elephant shark MR.   Interestingly, neither progesterone nor 19norProgesterone activated elephant shark GR, but these steroids did activate elephant shark MR, as previously reported (*36*).

**Effect of corticosteroids and progesterone on transcriptional activation of truncated elephant shark GR and MR.**

We investigated the role of the NTD in the response of elephant shark GR and MR to corticosteroids and progestins at $10^{-8}$ M, by constructing truncated elephant shark GR and MR, in which the NTD was deleted (Figure 3C, D).   Figure 3C shows that activation of truncated elephant shark GR by $10^{-8}$ M corticosterone decreased by over 90%, compared to full-length elephant shark GR (Figure 3A), and there was no significant activation of truncated GR by $10^{-8}$ M of either cortisol or 11-deoxycorticosterone, both of which activated full-length GR.   Progesterone did not activate truncated elephant shark GR.   In contrast, $10^{-8}$ M corticosteroids and progesterone had similar levels of activation of truncated elephant shark MR (Figure 3D) and full-length MR (Figure3B), indicating that there is a major difference in the role of the NTD in transcriptional activation of elephant shark GR and MR.

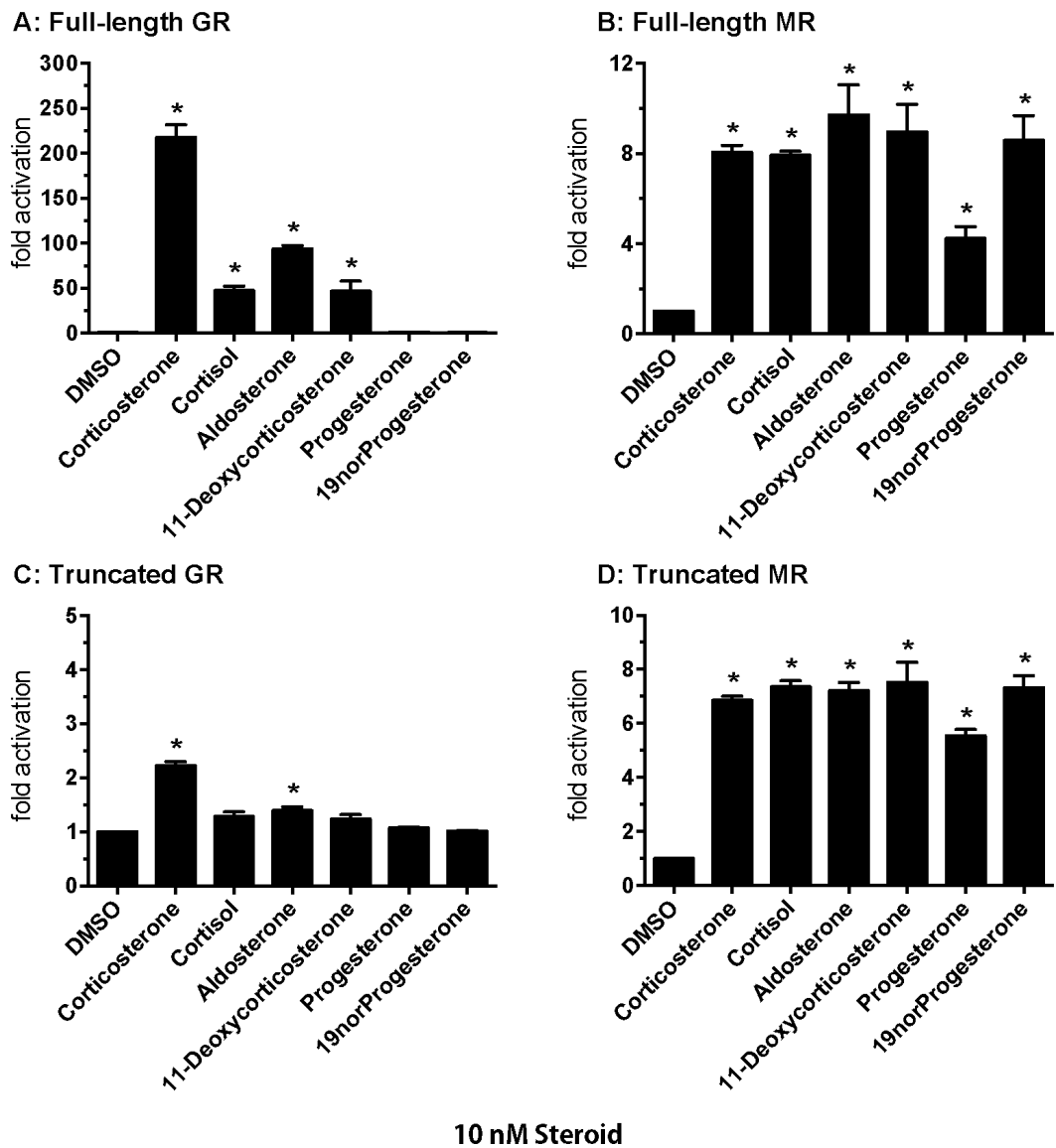

**Fig. 3. Ligand specificity of full-length and truncated elephant shark GR and MR.** Plasmids for full-length elephant shark GR or MR or truncated elephant GR or MR, containing the DBD (C domain), D domain and LBD (E domain), were expressed in HEK293 cells with an MMTV-luciferase reporter. Transfected cells were treated with either 10 nM cortisol, corticosterone, aldosterone, 11-deoxycorticosterone, progesterone, 19norProgesterone or vehicle alone (DMSO). Results are expressed as means ± SEM, n=3. Y-axis indicates fold-activation compared to the activity of control vector with vehicle (DMSO) alone as 1. A. Full-length GR. B. Full-length MR. C. Truncated GR. D. Truncated MR.

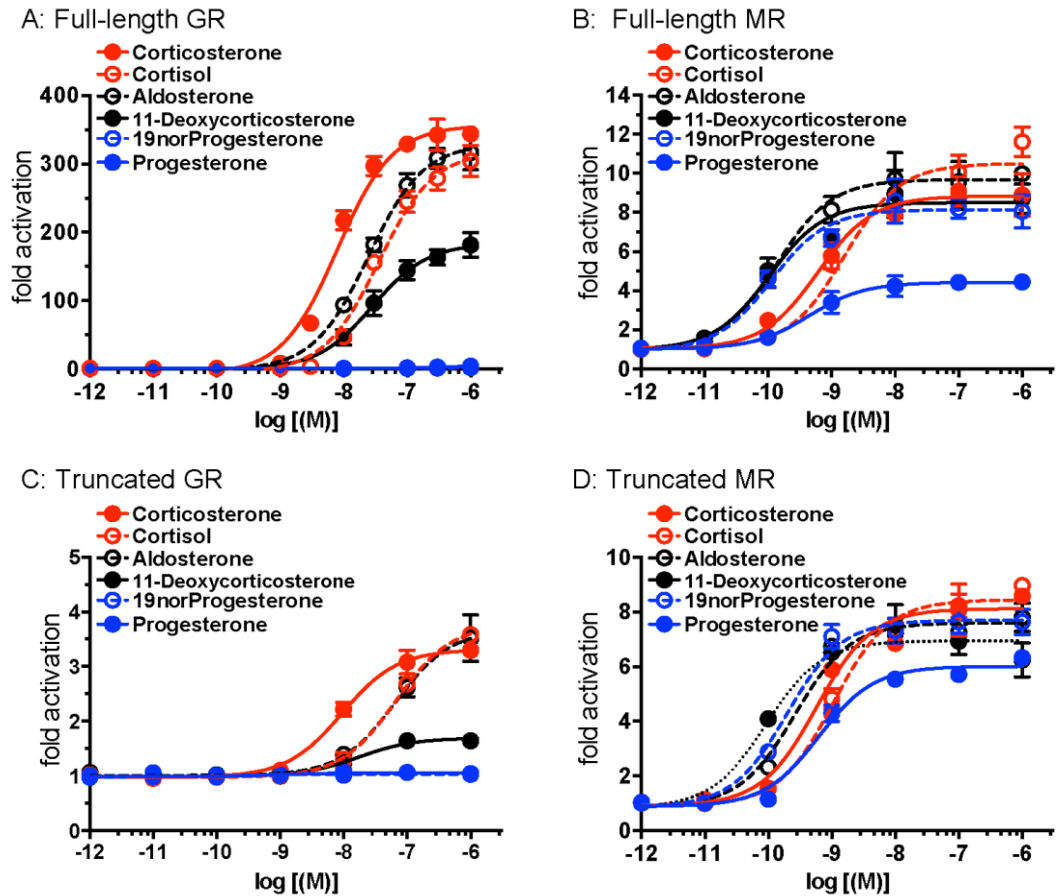

**Fig. 4. Concentration-dependent transcriptional activation by corticosteroids and progesterone of full length and truncated elephant shark GR and MR.** Plasmids for full-length elephant shark GR or MR or truncated elephant GR or MR, were expressed in HEK293 cells with an MMTV-luciferase reporter. Cells were treated with increasing concentrations of either cortisol, corticosterone, aldosterone, 11-deoxycorticosterone, progesterone, 19norProgesterone or vehicle alone (DMSO). Results are expressed as means ± SEM, n=3. Y-axis indicates fold-activation compared to the activity of control vector with vehicle (DMSO) alone as 1. A. Full-length GR. B. Full-length MR. C. Truncated GR. D. Truncated MR.

Next, we determined the concentration dependence of transcriptional activation by these corticosteroids of full-length and truncated elephant shark GR (Figure 4A, B) and MR (Figure 4C, D). This data was used to calculate EC50 values for steroid activation of full-length and truncated elephant shark GR and MR, which are shown in Table 1. Corticosterone and cortisol, two physiological glucocorticoids in mammals, had EC50s of 7.9 nM and 35 nM, respectively, for full-length elephant shark GR. For comparison, the EC50s of corticosterone and cortisol are 0.61 nM and 1.6 nM, respectively, for full-length elephant shark MR (Table 1) (*36*).   However, aldosterone and 11-deoxycorticosterone, two physiological mineralocorticoids in mammals, had EC50s of 24 nM and 28 nM respectively, for elephant shark GR, indicating that elephant shark GR retains some properties of its MR paralog.   For comparison, the EC50s of aldosterone and 11-deoxycorticosterone are 0.14 nM and 0.1 nM, respectively, for full-length elephant shark MR (Table 1) (*36*).   EC50s of corticosterone, cortisol and aldosterone for truncated GR were 9.4 nM, 66 nM and 65 nM respectively, while all four corticosteroids, progesterone and 19norProgesterone activated truncated MR with EC50s that were similar or lower than that of full-length MR.

Table 1 EC50s for Corticosteroid and Progestin Activation of Full-length and Truncated Elephant Shark GR and MR and GR-MR Chimeras

| Elephant Shark | Cortisol | Corticosterone | Aldosterone | 11-Deoxycorticosterone |
|---|---|---|---|---|
| GR-Full Length | 35 nM | 7.9 nM | 24 nM | 28 nM |
| GR-CDE (Truncated) | 66 nM | 9.4 nM | 65 nM | 17 nM |
| MR-AB+GR-CDE | 93 nM | 13 nM | 43 nM | - |
|  |  |  |  |  |
| MR-Full Length | 1.6 nM | 0.61 nM | 0.14 nM | 0.1 nM |
| MR-CDE (Truncated) | 1.1 nM | 0.58 nM | 0.26 nM | 0.09 nM |
| GR-AB+MR-CDE | 0.3 nM | 0.1 nM | 0.05 nM | 0.03 nM |

| Elephant Shark | Progesterone | 19norProgesterone |
|---|---|---|
| GR Full Length | - | - |
| GR-CDE (Truncated) | - | - |
| MR-AB+GR-CDE | - | - |
|  |  |  |
| MR-Full Length | 0.45 nM | 0.11 nM |
| MR-CDE (Truncated) | 0.63 nM | 0.2 nM |
| GR-AB+MR-CDE | 0.2 nM | 0.04 nM |

The dependence of the level of cortisol activation on the NTD in elephant shark GR is similar to that of human GR, in which deletion of the NTD reduces cortisol activation by over 90% (*17, 24*).   These results indicate that allosteric signaling between the NTD and DBD-LBD in elephant shark GR is critical for its response to corticosteroids, in contrast to elephant shark MR, in which truncated MR and full-length elephant shark MR have similar levels of activation by corticosteroids and progesterone.

**Progesterone binds to, but does not activate, full-length elephant shark GR.**
Neither progesterone nor19norProgesterone activated transcription by elephant shark GR (Figure 3A), although these steroids are transcriptional activators of elephant shark MR (Figure 3B) (*36*). However, unexpectedly, we find that progesterone and 19norProgesterone inhibit activation of elephant shark GR by 10 nM corticosterone (Figure 5A) indicating that elephant shark GR recognizes progesterone. A parallel study showed that neither progesterone nor 19norProgesterone inhibited activation of human GR by 10 nM cortisol (Figure 5B).

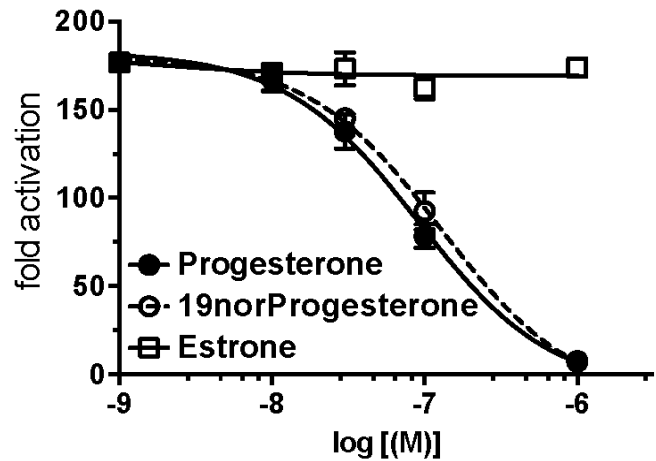

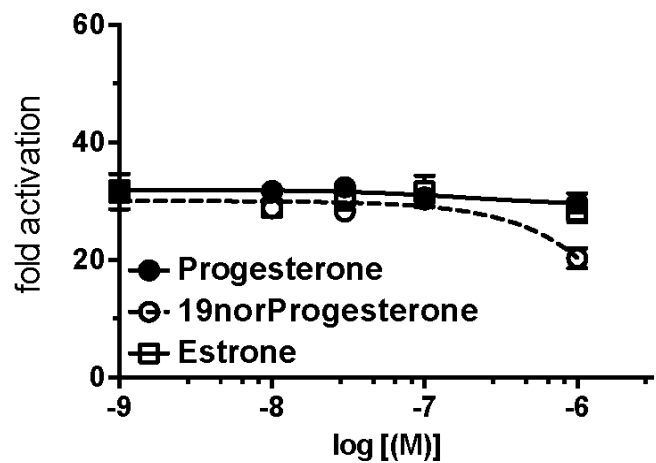

**Fig. 5. Progesterone inhibits corticosterone activation of full-length elephant shark GR.** Plasmids encoding full length elephant GR or human GR were expressed in HEK293 cells. Cells with elephant shark GR and human GR were treated with 10 nM corticosterone and cortisol respectively and increasing concentrations of either progesterone, 19norProgesterone or estrone. Y-axis indicates fold-activation compared to the activity of control vector with vehicle (DMSO) alone as 1.
A. Elephant shark GR. B. Human GR.

**Corticosteroid activation of chimeras in which the NTD is swapped between elephant shark GR and MR.**

To investigate further the evolution of transcriptional activity of the NTD in elephant shark GR and MR, we studied corticosteroid activation of elephant shark GR and MR chimeras, in which the GR NTD was fused to MR DBD-LBD and the MR NTD was fused to GR DBD-LBD (Figure 6).    The NTDs of the GR and MR had dominant effects on transcription of the DBD-LBD in each chimera (Figure 6, Table 1). Thus, in the GR NTD-MR DBD-LBD chimera (GR NTD fused to MR DBD-LBD), cortisol and corticosterone increased activation by over 30-fold, compared to full length elephant shark MR (Figures 3B, 4B and 6B).    Moreover, the GR NTD-MR DBD-LBD chimera also had a higher level of activation in the presence of progesterone and 19norProgestrone.    In addition, the EC50s for corticosteroids and progestin were lower in the GR NTD-MR DBD-LBD chimera indicating that the NTD affects the affinity of steroids for the chimera, as well as the level of transcriptional activation (Table 1).

In contrast, the MR NTD reduced corticosteroid activation of the MR NTD-GR DBD-LBD chimera by over 90% compared to full-length GR (Figures 3A, 4A and 6A). Moreover, the EC50 for cortisol increased over 2-fold, and an EC50 for 11-deoxycorticosterone was too low to be calculated (Table 1).    These data indicate that increased activation of the GR by the NTD in elephant shark GR evolved after it diverged from the MR.

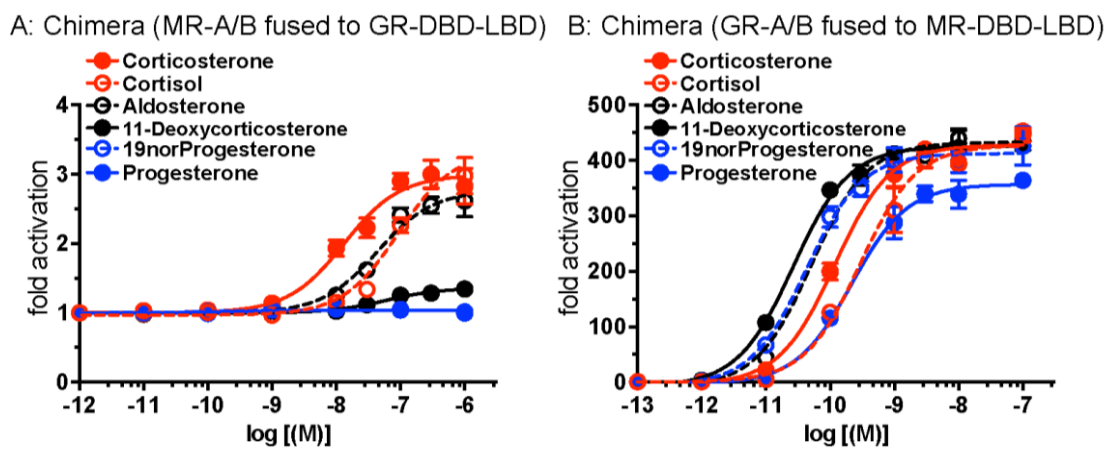

**Fig. 6. Concentration-dependent transcriptional activation by corticosteroids of chimeras of GR and MR.**    Plasmids for chimeric GR or MR were expressed in HEK293 cells with an MMTV-luciferase reporter.    Cells were treated with increasing concentrations of either cortisol, corticosterone, aldosterone, 11-deoxycorticosterone,

progesterone, 19norProgesterone or vehicle alone (DMSO).   Results are expressed as means ± SEM, n=3.   Y-axis indicates fold-activation compared to the activity of control vector with vehicle (DMSO) alone as 1.   A. Chimera of MR-A/B fused to GR-DBD-LBD.   B. Chimera of GR-A/B fused to MR-DBD-LBD.

**Discussion**

Sequence analysis revealed that the MR and GR are kin (*9*).   A distinct GR and MR, first appear in cartilaginous fish, which occupy a key position at an ancestral node from which ray-finned fish and terrestrial vertebrates diverged (*14, 38*).   To investigate early events in the evolution of steroid specificity for the GR during its divergence from the MR, we studied corticosteroid activation of full-length elephant shark GR, which we compared to similar experiments with elephant shark MR (*36*).   Because the NTD has a strong activation function for human GR (*17, 24*), we also studied elephant shark GR, in which the NTD was deleted.   After finding that corticosterone activation of this truncated GR declined by over 90% from the activity of full-length elephant shark GR, we investigated the evolution of the NTD in chimeras of elephant shark GR and MR, in which their NTDs were swapped.

Overall, we find that the response to corticosteroids by elephant shark GR has some similarities and differences with human GR (*17, 19, 22, 25*), as well as with elephant shark MR (*36*).   In comparison to corticosteroid activation of elephant shark MR, corticosteroids have higher EC50s for elephant shark GR, indicating reduced sensitivity to corticosteroids [Table 1].   Interestingly, the level of transcriptional activation by corticosteroids of elephant shark GR is over 10-fold higher than for elephant shark MR (Figure 3A, B).   In this respect, elephant shark GR and MR are similar to human GR and human MR in which cortisol has a higher EC50, as well as a higher level of transcriptional activation for human GR, compared to human MR (*17, 19, 39*).

Activation of elephant shark GR by aldosterone is a notable difference with human GR.   Although 10 nM aldosterone has little activity for human GR (Figure 3A) (*19, 25*), aldosterone and cortisol have a similar level of activation of elephant shark GR. Thus, while elephant GR evolved to respond to the physiological glucocorticoids in humans and other terrestrial vertebrates, elephant shark GR retains some activities of elephant shark MR.   We conclude that elephant shark GR is transitional from the MR in specificity for glucocorticoids and mineralocorticoids.

A novel property of elephant shark MR is that it is activated by progesterone and 19norProgesterone (*36*).   Neither steroid activates elephant shark GR.   Unexpectedly, progesterone and 19norProgesterone inhibit activation of elephant shark GR by corticosterone (Figure 5A), which suggests that progestins may influence activation of elephant shark GR.  Neither progesterone nor 19norProgesterone inhibit activation of human GR by cortisol (Figure 5B).  The timing of the loss of recognition of progestins by the GR needs further study.

The deletion of the NTD from elephant shark GR has similar effects on corticosteroid activation as found for human GR (*17, 24*), (Table 1, Figure 3D).  Thus, deletion of the NTD in human GR reduces cortisol and dexamethasone activity by over 90% (*17, 24*).  Similarly, we find that truncated elephant shark GR loses over 90% of activation by corticosteroids compared to full-length elephant shark GR.  In contrast, full-length and truncated elephant shark MRs have similar responses to corticosteroids (Figures 3 B, D, Table 1).

Experiments with chimeras in which the NTD of elephant shark GR and MR were swapped reveal that the NTD in the GR and MR have a dominant effect on transcriptional activation of each chimera.  Replacing the MR NTD with the GR NTD increases corticosteroid activation of the GR-NTD MR-LBD chimera compared to full-length MR, while activity of MR-NTD GR-LBD chimera is reduced to that of full-length MR.  This indicates that NTD stimulation of activation by corticosteroids of the GR arose early in vertebrate evolution.

The evolution of a distinct GR and MR in cartilaginous fish has provided an opportunity to investigate early events in the divergence of these two steroid receptors from their common ancestor.  Cartilaginous fish, including elephant shark, also contain a PR and AR, which diverged from a common ancestor.  The PR is a close relative of the MR.  Comparison of steroid activation of elephant shark PR and AR with each other and with the MR and GR could provide valuable insights into their evolution as important transcription factors in vertebrates.

**References**

1. D. W. Cain, J. A. Cidlowski, Immune regulation by glucocorticoids. *Nat Rev Immunol* **17**, 233-247 (2017).
2. E. R. de Kloet, From receptor balance to rational glucocorticoid therapy. *Endocrinology* **155**, 2754-2769 (2014).
3. K. L. Gross, J. A. Cidlowski, Tissue-specific glucocorticoid action: a family affair. *Trends Endocrinol Metab* **19**, 331-339 (2008).
4. R. G. Hunter *et al.*, Stress and corticosteroids regulate rat hippocampal mitochondrial DNA gene expression via the glucocorticoid receptor. *Proc Natl*

*Acad Sci U S A* **113**, 9099-9104 (2016).

5. E. R. Weikum, M. T. Knuesel, E. A. Ortlund, K. R. Yamamoto, Glucocorticoid receptor control of transcription: precision and plasticity via allostery. *Nature reviews. Molecular cell biology* **18**, 159-174 (2017).
6. M. E. Baker, D. R. Nelson, R. A. Studer, Origin of the response to adrenal and sex steroids: Roles of promiscuity and co-evolution of enzymes and steroid receptors. *J Steroid Biochem Mol Biol* **151**, 12-24 (2015).
7. J. T. Bridgham, S. M. Carroll, J. W. Thornton, Evolution of hormone-receptor complexity by molecular exploitation. *Science* **312**, 97-101 (2006).
8. R. M. Evans, The steroid and thyroid hormone receptor superfamily. *Science* **240**, 889-895 (1988).
9. J. L. Arriza *et al.*, Cloning of human mineralocorticoid receptor complementary DNA: structural and functional kinship with the glucocorticoid receptor. *Science* **237**, 268-275 (1987).
10. M. E. Baker, J. W. Funder, S. R. Kattoula, Evolution of hormone selectivity in glucocorticoid and mineralocorticoid receptors. *J Steroid Biochem Mol Biol* **137**, 57-70 (2013).
11. M. E. Baker, Y. Katsu, 30 YEARS OF THE MINERALOCORTICOID RECEPTOR: Evolution of the mineralocorticoid receptor: sequence, structure and function. *J Endocrinol* **234**, T1-T16 (2017).
12. K. S. Kassahn, M. A. Ragan, J. W. Funder, Mineralocorticoid receptors: evolutionary and pathophysiological considerations. *Endocrinology* **152**, 1883-1890 (2011).
13. B. C. Rossier, M. E. Baker, R. A. Studer, Epithelial sodium transport and its control by aldosterone: the story of our internal environment revisited. *Physiological reviews* **95**, 297-340 (2015).
14. B. Venkatesh *et al.*, Elephant shark genome provides unique insights into gnathostome evolution. *Nature* **505**, 174-179 (2014).
15. C. Hellal-Levy *et al.*, Specific hydroxylations determine selective corticosteroid recognition by human glucocorticoid and mineralocorticoid receptors. *FEBS Lett* **464**, 9-13 (1999).
16. Y. Katsu, K. Oka, M. E. Baker, Evolution of human, chicken, alligator, frog, and zebrafish mineralocorticoid receptors: Allosteric influence on steroid specificity. *Sci Signal* **11**,   (2018).
17. R. Rupprecht *et al.*, Transactivation and synergistic properties of the mineralocorticoid receptor: relationship to the glucocorticoid receptor.

*Molecular endocrinology (Baltimore, Md )* **7**, 597-603 (1993).

18. M. Lombes, S. Kenouch, A. Souque, N. Farman, M. E. Rafestin-Oblin, The mineralocorticoid receptor discriminates aldosterone from glucocorticoids independently of the 11 beta-hydroxysteroid dehydrogenase. *Endocrinology* **135**, 834-840 (1994).

19. J. L. Arriza, R. B. Simerly, L. W. Swanson, R. M. Evans, The neuronal mineralocorticoid receptor as a mediator of glucocorticoid response. *Neuron* **1**, 887-900 (1988).

20. R. Rupprecht *et al.*, Pharmacological and functional characterization of human mineralocorticoid and glucocorticoid receptor ligands. *Eur J Pharmacol* **247**, 145-154 (1993).

21. K. L. Gross, N. Z. Lu, J. A. Cidlowski, Molecular mechanisms regulating glucocorticoid sensitivity and resistance. *Molecular and cellular endocrinology* **300**, 7-16 (2009).

22. K. Oka *et al.*, Allosteric role of the amino-terminal A/B domain on corticosteroid transactivation of gar and human glucocorticoid receptors. *J Steroid Biochem Mol Biol* **154**, 112-119 (2015).

23. S. Cho, J. A. Blackford, Jr., S. S. Simons, Jr., Role of activation function domain-1, DNA binding, and coactivator GRIP1 in the expression of partial agonist activity of glucocorticoid receptor-antagonist complexes. *Biochemistry* **44**, 3547-3561 (2005).

24. S. M. Hollenberg, V. Giguere, P. Segui, R. M. Evans, Colocalization of DNA-binding and transcriptional activation functions in the human glucocorticoid receptor. *Cell* **49**, 39-46 (1987).

25. Y. Katsu, S. Kohno, K. Oka, M. E. Baker, Evolution of corticosteroid specificity for human, chicken, alligator and frog glucocorticoid receptors. *Steroids* **113**, 38-45 (2016).

26. R. Miesfeld, P. J. Godowski, B. A. Maler, K. R. Yamamoto, Glucocorticoid receptor mutants that define a small region sufficient for enhancer activation. *Science* **236**, 423-427 (1987).

27. A. Christopoulos *et al.*, International union of basic and clinical pharmacology. XC. multisite pharmacology: recommendations for the nomenclature of receptor allosterism and allosteric ligands. *Pharmacological reviews* **66**, 918-947 (2014).

28. S. H. Khan *et al.*, Binding of the N-terminal region of coactivator TIF2 to the intrinsically disordered AF1 domain of the glucocorticoid receptor is accompanied by conformational reorganizations. *The Journal of biological*

*chemistry* **287**, 44546-44560 (2012).

29. A. Sturm, J. E. Bron, D. M. Green, N. R. Bury, Mapping of AF1 transactivation domains in duplicated rainbow trout glucocorticoid receptors. *J Mol Endocrinol* **45**, 391-404 (2010).

30. A. Sturm, L. Colliar, M. J. Leaver, N. R. Bury, Molecular determinants of hormone sensitivity in rainbow trout glucocorticoid receptors 1 and 2. *Molecular and cellular endocrinology* **333**, 181-189 (2011).

31. P. J. Fuller, Y. Yao, J. Yang, M. J. Young, Mechanisms of ligand specificity of the mineralocorticoid receptor. *J Endocrinol* **213**, 15-24 (2012).

32. U. A. Hawkins, E. P. Gomez-Sanchez, C. M. Gomez-Sanchez, C. E. Gomez-Sanchez, The ubiquitous mineralocorticoid receptor: clinical implications. *Curr Hypertens Rep* **14**, 573-580 (2012).

33. O. M. Conneely, B. Mulac-Jericevic, F. DeMayo, J. P. Lydon, B. W. O'Malley, Reproductive functions of progesterone receptors. *Recent Prog Horm Res* **57**, 339-355 (2002).

34. S. Publicover, C. Barratt, Reproductive biology: Progesterone's gateway into sperm. *Nature* **471**, 313-314 (2011).

35. Dobzhansky.T, Nothing in Biology Makes Sense except in the Light of Evolution. *Am Biol Teach* **35**, 125-129 (1973).

36. Y. Katsu *et al.*, Transcriptional activation of elephant shark mineralocorticoid receptor by corticosteroids, progesterone, and spironolactone. *Sci Signal* **12**, (2019).

37. S. M. Carroll, J. T. Bridgham, J. W. Thornton, Evolution of hormone signaling in elasmobranchs by exploitation of promiscuous receptors. *Molecular biology and evolution* **25**, 2643-2652 (2008).

38. J. G. Inoue *et al.*, Evolutionary origin and phylogeny of the modern holocephalans (Chondrichthyes: Chimaeriformes): a mitogenomic perspective. *Molecular biology and evolution* **27**, 2576-2586 (2010).

39. P. Kiilerich *et al.*, Interaction between the trout mineralocorticoid and glucocorticoid receptors in vitro. *J Mol Endocrinol* **55**, 55-68 (2015).

40. K. Oka *et al.*, Molecular cloning and characterization of the corticoid receptors from the American alligator. *Molecular and cellular endocrinology* **365**, 153-161 (2013).

41. Y. Katsu, K. Kubokawa, H. Urushitani, T. Iguchi, Estrogen-dependent transactivation of amphioxus steroid hormone receptor via both estrogen and androgen response elements. *Endocrinology* **151**, 639-648 (2010).

**Acknowledgments:**

**Funding**: This work was supported in part by Grants-in-Aid for Scientific Research 19K0673409 (YK) from the Ministry of Education, Culture, Sports, Science and Technology of Japan.   M.E.B. was supported by Research fund #3096.

**Author contributions**: Y.K. and XXX carried out the research.   Y.K. and M.E.B. conceived and designed the experiments.   MEB wrote the paper.   All authors gave final approval for publication.

**Competing interests**: We have no competing interests.

**Data and materials availability**: not applicable.

**Supplementary Material**

**Materials and Methods**

**Chemical reagents.** Aldosterone, cortisol, corticosterone, 11-deoxycorticosterone, , progesterone and 19norProgesterone were purchased from Sigma-Aldrich.   For the reporter gene assays, all hormones were dissolved in dimethylsulfoxide (DMSO) and the final concentration of DMSO in the culture medium did not exceed 0.1%.

**Construction of plasmid vectors** The full-coding regions and DBD-LBD domains of the GR and MR from *C. milii* were amplified by PCR with KOD DNA polymerase (TOYOBO Biochemicals, Osaka, Japan).   The PCR products were gel-purified and ligated into pcDNA3.1 vector (Invitrogen) (*40*).

**Transactivation Assay and Statistical Methods.** Hek293 were used in the reporter gene assay.   Transfection and reporter assays were carried out as described previously (*40, 41*).   All transfections were performed at least three times, employing triplicate sample points in each experiment.   The values shown are mean ± SEM from three separate experiments.   Data from studies of the effect of different corticosteroid concentrations on transcriptional activation of elephant shark GR and MR were used to calculate EC50s for steroid activation of elephant shark GR and MR using GraphPad Prism.   Comparisons between two groups were performed using *t*-test, and all multi-group comparisons were performed using one-way ANOVA followed by Bonferroni test.   $P < 0.05$ was considered statistically significant.

**Genbank Accessions for Domain Comparison.** Elephant shark GR (XP_007899521), human GR (BAH02307), chicken GR (ABB05045), *X. laevis* GR (NP_001081531), Zebrafish GR (NP_001018547), elephant shark MR (XP_007902220), human MR (NP_000892), zebrafish MR (NP_001093873).